\documentclass[aps,prl,twocolumn,superscriptaddress,nofootinbib,amsmath,amssymb,floatfix]{revtex4}  
\usepackage{graphicx}  
\usepackage{dcolumn}   
\usepackage{bm,color}        
\usepackage{amssymb}   

\newcommand{\be}{\begin{equation}}
\newcommand{\ee}{\end{equation}}

\newcommand{\Mpl}{M_{\textrm{Pl}}}

\newcommand{\rd}{\mathrm{d}}
\newcommand{\rmd}{\mathrm{d}}

\newcommand{\cT}{c_T}

\begin{document}

\title{Resilience of the standard predictions for primordial tensor modes}
\author{Paolo Creminelli} 
\address{Abdus Salam International Centre for Theoretical Physics\\ Strada Costiera 11, 34151, Trieste, Italy  
         }
\author{J\'er\^ome Gleyzes}
\address{CEA, Institut de Physique Th{\'e}orique,
         91191 Gif-sur-Yvette c{\'e}dex,  France \\
         CNRS, Unit{\'e} de recherche associ{\'e}e-2306, 91191 Gif-sur-Yvette c{\'e}dex, France}
\address{Universit\'e Paris Sud, 15 rue George Cl{\'e}menceau, 91405,  Orsay, France}
\author{Jorge Nore\~na}
\address{Department of Theoretical Physics and Center for Astroparticle Physics (CAP)\\24 quai E.~Ansermet, 1211 Geneva 4, Switzerland
         }
\author{Filippo Vernizzi}
\address{CEA, Institut de Physique Th{\'e}orique,
         91191 Gif-sur-Yvette c{\'e}dex,  France \\
         CNRS, Unit{\'e} de recherche associ{\'e}e-2306, 91191 Gif-sur-Yvette c{\'e}dex, France}
        \date{\today}

\begin{abstract}
We show that the prediction for the primordial tensor power spectrum cannot be modified at leading order in derivatives. Indeed, one can always set to unity the speed of propagation of gravitational waves during inflation by a suitable disformal transformation of the metric, while a conformal one can make the Planck mass time-independent. Therefore, the tensor amplitude unambiguously fixes the energy scale of inflation. Using the Effective Field Theory of Inflation, we check that predictions are independent of the choice of frame, as expected.  The first corrections to the standard prediction come from two parity violating operators with three derivatives. Also the correlator $\langle\gamma\gamma\gamma\rangle$ is standard and only receives higher derivative corrections. These results hold also in multifield models of inflation and in alternatives to inflation and make the connection between a (quasi) scale-invariant tensor spectrum and inflation completely robust.
\end{abstract}

\maketitle

{\em Introduction -}
We are entering an exciting period for primordial gravitational waves, since BICEP2 \cite{Ade:2014xna} has shown that the experimental sensitivity to $B$-modes is now at the level of an interesting regime for tensors, provided foreground contamination is under control.  The importance of primordial tensor modes lies in their robustness: while scalar perturbations are sensitive to many details (the shape of the potential, the speed of propagation of scalar fluctuations $c_s$, the number of fields and their conversion to adiabatic perturbations) and can also be viably produced in non-inflationary models, tensor modes are much more model independent. In this Letter we strengthen this robustness, showing that one cannot change the tensor quadratic and cubic action at leading order in derivatives. Since the inflaton defines a preferred frame, the time and spatial kinetic term of gravitons can have in general different time-dependent coefficients. However, without loss of generality, one can always make the graviton speed equal to unity by doing a suitable disformal transformation. A conformal transformation can then remove any time dependence of the overall normalization of the action,~i.e., the Planck mass, so that the dynamics of gravitons is completely standard.

\vspace{0.1cm}
{\em Disformal vs Einstein frame -} 
We work here with the (single-field) Effective Field Theory of Inflation \cite{Creminelli:2006xe,Cheung:2007st} and we will comment on generalizations later.
Working in unitary gauge, where the inflaton perturbations are set to zero, the speed of gravitons can be changed by the operator $\delta K_{\mu\nu} \delta K^{\mu\nu}$, where $\delta K_{\mu \nu} $ is the perturbation of the extrinsic curvature of the spatial slices, $ K_{\mu \nu} $ \cite{Cheung:2007st,Gleyzes:2013ooa,Noumi:2014zqa}. 
This kind of modifications arises when considering higher derivative operators for the inflaton, such as in Horndeski theories  \cite{Kobayashi:2011nu}.
We are free to subtract   $\delta K^2$, which  at  quadratic order contains only scalars. As shown below, the combination $\delta K_{\mu\nu} \delta K^{\mu\nu} - \delta K^2$ does not change the  sound speed of scalar fluctuations.  
Thus, we consider the action
\be
\begin{split}
\label{eq:NaiveAction}
S = \int  \mathrm{d}^4 x&\sqrt{-g} \frac{\Mpl^2}{2} \Big[R  - 2  \big( \dot{H} + 3 H^2\big) + 2  \dot{H} g^{00}\\ 
 &- \big(1 - \cT^{-2} (t)\big)\big(\delta K_{\mu\nu} \delta K^{\mu\nu} - \delta K^2\big) \Big]\,,
\end{split}
\ee
where $H \equiv \dot a/a$ is the Hubble rate and the first line describes a minimal slow-roll model \cite{Cheung:2007st}. 

We will use the usual ADM decomposition,
 \be
 \mathrm{d} s^2 = - N^2 \mathrm{d} t^2 + h_{ij} (N^i \mathrm{d} t + \mathrm{d} x^i) (N^j \mathrm{d} t + \mathrm{d} x^j) \;,
 \ee
 and describe scalar and tensor perturbations as \cite{Maldacena:2002vr}
 \be
\label{hij}
h_{ij} = a^2 e^{2 \zeta} (e^\gamma)_{ij} \;, \qquad \gamma_{ii}=0=\partial_i \gamma_{ij} \;.
\ee
In these variables the extrinsic curvature is given by
\be
\label{eq:K}
K_{ij} = \frac{1}{2N} \big( \dot h_{ij} - \nabla_i N_j - \nabla_j N_i \big)\;.
\ee
The coefficient in the second line of eq.~\eqref{eq:NaiveAction} is chosen such that the tensor quadratic action  reads
\be
\label{qa_gamma}
S_{\gamma \gamma} = \frac{\Mpl^2}{8}  \int \mathrm{d}^4 x a^3 \cT^{-2} \bigg[ \dot{\gamma}_{ij}^2 - \cT^2 \frac{(\partial_k\gamma_{ij})^2}{a^2} \bigg]\;.
\ee
The second line of \eqref{eq:NaiveAction} modifies the time kinetic term of gravitons; the only other way to change tensor modes at quadratic order is to modify the spatial kinetic term with the operator $^{(3)}\!R$, the 3d Ricci tensor. The two choices are related by the Gauss-Codazzi identity,
\be
\label{GC}
R = {}^{(3)}\! R  + K_{\mu \nu} K^{\mu \nu}- K^2 + 2 \nabla_\mu (K n^\mu - n^\rho \nabla_\rho n^\mu ) \;,
\ee
where $n^\mu$ is the unit vector perpendicular to the surfaces of constant time.

The main point of this paper is that it is possible to set to one the speed of propagation of gravitons in action \eqref{qa_gamma} by a  proper redefinition of the metric.
Metric transformations that change the light-cone are known as disformal transformations \cite{Bekenstein:1992pj}, so that we denote the metric used to write eq.~\eqref{eq:NaiveAction} as the  disformal metric. We first perform  a disformal transformation which leaves the spatial  metric in unitary gauge unchanged,\footnote{In terms of the inflaton field $\phi$, the new metric reads $g_{\mu\nu} \mapsto  g_{\mu\nu} - (1- \cT^2 ) \partial_\mu\phi\partial_\nu\phi /(\partial \phi)^2$.}\,\footnote{A similar transformation was also employed for instance in \cite{Germani:2011bc} to set an action with modified graviton sound speed in the standard Einstein-frame form.}
\begin{equation}
\label{eq:disformal}
g_{\mu\nu} \mapsto  g_{\mu\nu} + (1- \cT^2(t))n_\mu n_\nu \;.
\end{equation}
This transformation does not affect $N^i$ and $h_{ij}$ while $N \mapsto \cT N$. Thus $K_{ij} \mapsto K_{ij}/\cT$, while $ {}^{(3)}\! R$ is not changed. In this way the relative coefficient between the time and the spatial kinetic term of gravitons can be set to one and combined to give the 4d Ricci scalar through \eqref{GC}. However, the normalization of the Einstein-Hilbert term is now non-standard and given by $\frac12 \Mpl^2 R/\cT(t) $. This can be cast in the standard form by going to the Einstein frame with a conformal transformation of the metric,
\be
g_{\mu \nu} \mapsto \cT^{-1} (t) \, g_{\mu \nu}\;.
\ee
Notice that in doing the disformal and conformal transformations 
the FLRW line element becomes
 $\mathrm{d} \tilde s^2 = \cT^{-1}[-\cT^2 \mathrm{d}t^2 + a^2 \mathrm{d} \vec x^2]$.
It is thus convenient to redefine the  time coordinate and the scale factor as 
\be
\label{time_red}
\tilde t \equiv \int \cT^{1/2}(t) \mathrm{d} t\;, \qquad \tilde a (\tilde t) \equiv \cT^{-1/2} a(t) \;.
\ee

Under this combined set of transformations the components of the metric in Einstein frame read
$\tilde g^{00} = g^{00}$ ($g^{00}= - 1/N^2$), $\tilde N^i = \cT^{1/2} N^i$ and $\tilde h_{ij} = \cT^{-1} h_{ij}$. Using these relations it is straightforward to compute the Einstein-frame action,
\begin{align}
S  &=  \int  \mathrm{d} \tilde t \mathrm{d}^3 x   \sqrt{-  \tilde g} \frac{\Mpl^2}{2 } \bigg\{  \tilde R -2 \big( \dot{\tilde H} + 3  \tilde H^2 \big) + 2 \dot{{\tilde H}} \tilde  g^{00}
\nonumber \\ 
&+\bigg[ 2 \big(1-\cT^2\big) \dot{{\tilde H}}- \frac32 \alpha^2 - \cT^2 \bigg( \dot \alpha +  \tilde H \alpha + \frac12 \alpha^2 \bigg) \bigg] \nonumber \\& \times \Big( 1 - \sqrt{- \tilde g^{00} } \Big)^2  
 +   2 \alpha \,\delta \tilde K \Big( 1 - \sqrt{- \tilde g^{00} } \Big) \bigg\}\,,  \label{eq:Action}
\end{align}
where $\alpha \equiv \dot \cT/\cT$. Here and in the action above time derivatives are with respect to $\tilde t$. The last term in the action is obtained when using the Gauss-Codazzi identity to combine 3d quantities to form the 4d Ricci scalar, by integrating by parts the last term of \eqref{GC}. The first line has the expected dependence on the background evolution in Einstein frame, while the rest starts quadratic in the perturbations. In this frame, the kinetic term of gravitons is the standard one, given by the Einstein-Hilbert term.
 If $\alpha =0$ we just have a polynomial in $\tilde g^{00}+1$,  which describes an inflationary model with a Lagrangian of the form $P(\phi,(\partial\phi)^2)$. 
 
We stress that in doing disformal and conformal transformations one changes the way other particles are coupled to the metric; this however is immaterial, since it does not enter in the inflationary predictions. 

 \vspace{0.1cm}
{\em Frame independence of predictions -}
Since the definition of $\zeta$ and $\gamma_{ij}$ is the same in the disformal and Einstein frame, we expect all the inflationary predictions to remain unchanged, as we are now going to show.
We start by discussing the scalar fluctuations. 
It is important to note that in the disformal frame, for significant modifications of $\cT^2$, the coefficient in front of $\delta K_{\mu\nu} \delta K^{\mu\nu} - \delta K^2$  in action \eqref{eq:NaiveAction} is of order  $\Mpl^2$. Thus, one cannot rely on the decoupling limit when deriving predictions from this action.

As anticipated above, the operator in the second line of eq.~\eqref{eq:NaiveAction} does not contribute to scalar fluctuations up to quadratic order. 
Indeed, to fix $N$ we need the solution of the momentum constraint, which  is the same as in the standard $\cT=1$ case, i.e.~$ N = 1+\dot \zeta/H$ \cite{Maldacena:2002vr} (use for instance eq.~(74) of \cite{Piazza:2013coa}). Thus, from eq.~\eqref{eq:K} the scalar contributions  to $K_{ij}$ from $ N$ and $\dot h_{ij}$ cancel and we are left with those coming from $N^i$ which, in the combination that appears in eq.~\eqref{eq:NaiveAction}, only give a total derivative. 
Thus, the scalar sound speed in the disformal frame is $c_s=1$.

Since in the Einstein frame tensor modes propagate on the light-cone, we expect the scalar speed of propagation to be $\tilde c_s = 1/\cT$. For a constant $\cT$ ($\alpha=0$), this can be easily seen from the first term on the second line of action \eqref{eq:Action}. Indeed, introducing the scalar Goldstone boson $\tilde \pi$ associated with the breaking of time-diff invariance by the time transformation $\tilde t \mapsto \tilde t + \tilde \pi (\tilde t, \vec x)$, and expanding up to cubic order in the decoupling limit, the action becomes 
\be
\label{pi_action}
{\cal L} = \tilde a^3  {\Mpl^2 |\dot{ \tilde H} |} \cT^{2}  \bigg[ \dot {\tilde \pi}^2 - \cT^{-2} \frac{(\partial_i \tilde \pi)^2}{\tilde a^2}  - (1-\cT^{-2} ) \dot {\tilde \pi} \frac{(\partial_i \tilde \pi)^2}{ \tilde a^2}\bigg].
\ee
One can verify that $\tilde c_s = 1/\cT$, as expected, also when $\alpha \neq 0$ (use e.g.~eq.~(69) of \cite{Gubitosi:2012hu}).

Let us now check that the  spectrum of gravitational waves is the same when computed  in either frame. For the quadratic action \eqref{qa_gamma}, scale invariance is obtained for $a\,\cT^{-1/2} \int (\cT /a) dt   \simeq $ const.~(we do not assume $\cT$ slowly varying, see \cite{Khoury:2008wj}). Perturbations evolve with an effective scale factor $a \,\cT^{-1/2}$ so that the gravitational wave spectrum becomes 
\be
\label{Ef_GWPS}
\langle \gamma^s_{\vec k} \gamma^{s'}_{\vec k'} \rangle = (2 \pi)^3 \delta (\vec k + \vec k') \frac{1}{2 k^3}\frac{ (H-\alpha/2)^2}{\Mpl^2 \cT} \delta_{s s'} \;.
\ee
(The polarization tensors $\epsilon_{ij}^s$ are normalized as $\epsilon_{ij}^s \epsilon_{ij}^{s'}=4 \delta_{s s'}$ where $s,s' $ denote the helicity states.)
Using eq.~\eqref{time_red}, the Einstein frame Hubble rate is $\tilde H = \cT^{-1/2} (H-\alpha/2)$, implying that eq.~\eqref{Ef_GWPS} is simply the standard spectrum for gravitational waves with unit sound speed in Einstein frame. 
It is straightforward to verify, using again eq.~\eqref{time_red}, that also the scalar power spectrum is the same in both frames.

Given that the relation between the two frames does not involve the spatial coordinates, also the tilt of the tensor and scalar power spectra remains the same. For tensors, this is given by the usual formula $n_T = 2 \dot {\tilde H}/\tilde H^2$.
In the disformal frame, it is possible to obtain a blue tilt by a time varying $\cT$, keeping $\dot H < 0$. In this case one does not violate the Null Energy Condition (NEC) and, indeed, there is no sign of instability.
It is interesting to see how this  translates in the Einstein frame where a blue tilt requires $\dot{ \tilde H} > 0$.
One can check that the usual gradient instability associated with the violation of the NEC is cured by the last term of action \eqref{eq:Action}, as showed in \cite{Creminelli:2006xe}. For example, this operator arises in Galileon models that violate the NEC \cite{Creminelli:2010ba}.

We conclude that there is no loss of generality in assuming that gravitons have a standard kinetic term. In particular, this implies that the amplitude of tensor modes is fixed by the vacuum energy of inflation and that a blue spectrum of gravitational waves, $n_T >0$, requires $\dot {\tilde H} >0$, i.e.~a violation of the NEC  in Einstein frame. Moreover, the observation of an approximately scale-invariant tensor spectrum would imply an approximately time-independent $\tilde H$. While one can make a scale-invariant scalar power spectrum playing with a variable speed of sound $c_s$ and equation of state $\epsilon \equiv -\dot H /H^2$ \cite{Khoury:2008wj}, tensors are absolutely robust and sensitive only to $\tilde H$. It is worthwhile to stress that these conclusions do not change if we consider multifield models of inflation, or even alternatives to inflation. However, our conclusions do not apply to cases with a different symmetry structure, like solid inflation \cite{Gruzinov:2004ty} (in this case one can have $n_T>0$ with $\dot {\tilde H} <0$) or gauge-flation \cite{Maleknejad:2012fw}, or when tensors are produced not as vacuum fluctuations \cite{Cook:2011hg}.

 \vspace{0.1cm}
{\em Non-Gaussianity -}
We now show the equivalence between the two frames beyond linear order, taking $c_T$ time-independent for simplicity. We saw that in Einstein frame the scalar has a nontrivial sound speed $\tilde c_s = 1/\cT$. This implies a cubic interaction $\propto (1-\tilde c_s^{-2})$, as in eq.~\eqref{pi_action}. In the disformal frame this is not obvious, since the second line of action \eqref{eq:NaiveAction} does not contribute to the action of $\pi$ in the decoupling limit. However, as mentioned above, one cannot rely on this limit, but has to solve the constraints.
The linear Hamiltonian constraint fixes the scalar part of the shift. Crucially, this gets rescaled by a factor $\cT^2$ with respect to the standard case (use eq.~(75) of \cite{Piazza:2013coa}),
\be
\label{eq:psi}
\psi \equiv \partial^{-2} \partial_i N^i = - \cT^2  \frac{\zeta}{a^2H} +\chi \;, \quad \partial^2 \chi = \epsilon \cT^2  \dot \zeta\;.
\ee 
Using this solution, after several manipulations and integration by parts, one obtains that the leading interaction in the slow-roll limit, up to field redefinitions which die out on  super-Hubble scales, is
\be
\label{scalar_NG}
 {\cal L}_{\zeta \zeta \zeta} = a   \epsilon \big(1-\cT^2\big) \frac{\dot \zeta }H   (\partial_i \zeta)^2 \;,
\ee
which yields  $f_{\rm NL} \sim 1-\cT^2 = 1-\tilde c_s^{-2}$.

Let us now discuss cubic interactions involving gravitons. As already noticed in \cite{Maldacena:2011nz},  the second line of eq.~\eqref{eq:NaiveAction} does not contain cubic graviton vertices. Therefore, in both frames $\langle\gamma\gamma\gamma\rangle$ coincides with the minimal slow-roll result of  \cite{Maldacena:2002vr}. To study interactions involving two gravitons and one scalar we need to expand the action to cubic order and plug in the linear solutions to the constraints, i.e. $N=1+ \dot \zeta/H$ and eq.~\eqref{eq:psi}. After some manipulations and integrations by parts (see \cite{Maldacena:2002vr}) one obtains, at leading order in slow-roll,
\be
 {\cal L}_{\gamma \gamma \zeta}= \frac{\Mpl^2}{8}a^3  \cT^{-2}  \bigg[ \epsilon  \zeta \bigg( \dot\gamma_{ij}^2+ \cT^2 \frac{(\partial\gamma_{ij})^2}{a^2} \bigg)-2\dot\gamma_{ij}\partial\gamma_{ij}\partial \chi  \bigg] .
 \ee
In the Einstein frame the cubic interaction is standard (see eq.~(3.17) of \cite{Maldacena:2002vr}) except for a factor of $\cT^2$ in the solution for $\chi$ due to the scalar speed of sound (see eq.~(4.9) of \cite{Chen:2006nt}). Taking into account eq.~\eqref{time_red} and the different wavefunctions, one can check that $\langle\gamma\gamma\zeta\rangle$ computed in the two frames coincide. 
This correlator goes as $\langle\gamma\gamma\zeta\rangle \sim  \epsilon  \langle\zeta\zeta\rangle\langle\gamma\gamma\rangle$.\footnote{The cubic $\gamma \gamma \zeta$ action is suppressed by $\epsilon \zeta$ compared to the graviton kinetic term. This holds also for the term including $\chi$ in the limit $\tilde c_s \ll 1$ since, in the Einstein frame,
\be
\frac{{\cal L}_{\gamma\gamma\zeta}}{\Mpl^2} \supset \dot \gamma \partial \gamma \partial \chi \sim \epsilon \, \tilde c_s^{-2} \dot \gamma \partial \gamma \partial^{-1} \dot \zeta \sim \epsilon \, \dot \gamma \partial \gamma \frac{\partial}{\tilde H} \zeta \sim \epsilon \, \dot \gamma^2 \zeta \;,
\ee
where we used $\dot \zeta \sim \tilde c_s^2 \partial^2 \zeta/\tilde H$. Indeed, given the different dispersion relation, $\zeta$ is already frozen when tensor modes exit the Hubble radius. 
} This  differs from the result of \cite{Noumi:2014zqa} obtained in the decoupling limit. 
Finally, it is straightforward to verify that also the prediction for $\langle \gamma \zeta \zeta \rangle$ is  the same in the two frames and coincides with the minimal slow-roll model \cite{Maldacena:2002vr}.

\vspace{0.1cm} 
{\em Quadratic terms with three derivatives -}
We have seen that it is possible, without loss of generality, to cast the graviton kinetic term in the standard form. From now on we assume to be in Einstein frame and we drop the tildes. Notice that the operators $\dot\gamma_{ij}^2$ and $(\partial_l \gamma_{ij})^2$ are the only quadratic operators with two derivatives. Indeed, one could imagine a term with one time and one space derivative, in the parity violating combination $\varepsilon^{ijk} \partial_i \gamma_{jl} \dot \gamma_{lk}$, where $\varepsilon^{ijk} $ is the totally antisymmetric tensor. However, it is easy to see that this is a total derivative.

The first possible corrections to the tensor power spectrum come from terms with three derivatives. The combinations with an even number of spatial derivatives, $\dot\gamma_{ij} \ddot\gamma_{ij}$ and $\partial_l \gamma_{ij} \partial_l \dot\gamma_{ij}$, are total derivatives, so we are left to consider parity-violating terms with one or three spatial derivatives. There are two possible combinations,
\be
\label{Poddop}
\varepsilon^{ijk} \partial_i \dot\gamma_{jl} \dot \gamma_{lk}\;, \qquad \varepsilon^{ijk} \partial_i \partial_m \gamma_{jl} \partial_m \gamma_{lk}\;.
\ee
The first term comes from $ 4 \int \mathrm{d} ^4x \;\varepsilon^{0ijk} \nabla_i \delta K_{jl} \delta K_{lk}$.
The second term comes from the 3d Chern-Simons term,
\be
-4 \int \mathrm{d}^4x \;\varepsilon^{ijk} \left(\frac12 {}^3\Gamma^p_{iq}\partial_j {}^3\Gamma^q_{kp} +\frac13 {}^3\Gamma^p_{iq} {}^3\Gamma^q_{jr} {}^3\Gamma^r_{kp}\right) \;,
\ee 
where ${}^3\Gamma^i_{jk}$ are the Christoffel symbols of the 3d metric. The impact of these terms on primordial gravitational waves has been studied in the context of Horava-Lifschitz gravity in \cite{Takahashi:2009wc, Wang:2012fi}.\footnote{Parity violation in the context of inflation \cite{Lue:1998mq} is usually discussed in terms of the topological current
\be
K^\mu = 2 \varepsilon^{\mu\alpha\beta\gamma}  \left(\frac12 \Gamma^\sigma_{\alpha\nu}\partial_\beta \Gamma^\nu_{\gamma\sigma} +\frac13 \Gamma^\sigma_{\alpha\nu} \Gamma^\nu_{\beta\eta} \Gamma^\eta_{\gamma\sigma}\right) \;,
\ee
which satisfies
\be
\partial_\mu K^\mu = \frac14 \varepsilon^{\mu\nu\alpha\beta}  R^\sigma_{\ \rho\alpha\beta} R^\rho_{\ \sigma\mu\nu} \;.
\ee
It is easy to see that the operator $-2 \int \rd^4x \; K^0$ gives, at quadratic order in $\gamma$, the linear combination $\varepsilon^{ijk} \partial_i \dot\gamma_{jl} \dot \gamma_{lk} - \varepsilon^{ijk} \partial_i \partial_m \gamma_{jl} \partial_m \gamma_{lk}$. Notice, however, that in general the relative coefficient of the two operators in eq.~\eqref{Poddop} is not fixed by symmetry.}

It is easy to study the effect of the two 3-derivative operators on the power spectrum of tensor modes. The standard quadratic action is modified by the addition of 
\be
\label{parityviol}
- \frac{\Mpl^2}{8} \int \rd^4x \frac{1}{H \eta} \left[\frac{\alpha}{\Lambda} \varepsilon^{ijk}\partial_i \gamma_{jl}' \gamma_{lk}' + \frac{\beta}{\Lambda}\varepsilon^{ijk} \partial_i\partial_m\gamma_{jl}\partial_m\gamma_{lk}\right],
\ee
where a prime denotes the derivative with respect to the conformal time $\eta \equiv \int dt/a$, $\alpha$ and $\beta$ are dimensionless coefficients and $\Lambda$ is the scale that suppresses these higher dimension operators. We are going to assume an exact de Sitter background and take $\alpha$ and $\beta$, which could depend on time, to be approximately constant. In this limit the dilation isometry of de Sitter guarantees the spectrum to remain scale invariant also in the presence of the new operators. We are going to treat the corrections due to these terms perturbatively, i.e.~assume that the energy scale of the problem, the Hubble scale $H$, is small compared to $\Lambda$. The action \eqref{parityviol} violates parity and induces opposite corrections to the power spectrum of gravitons with opposite helicities. Indeed, the polarization tensors $\epsilon^{\pm}_{ij}$ of the two helicities satisfy $i k_l \, \varepsilon^{j l m} \epsilon^\pm_{im} = \pm k \,\epsilon^{\pm\;j}_i$. The interaction Hamiltonian ${\cal H}_{\rm int}$ in Fourier space is thus given by
\be
{\cal H}_{\rm int} = \pm\frac{\Mpl^2}{2 H \Lambda} \int  \frac{\rmd^3 k}{(2\pi)^3} \frac{ k}{\eta} \left[\alpha   \gamma_{\vec k}^{\pm}{}'  \gamma_{-\vec k}^{\pm}{}' + \beta k^2 \gamma_{\vec k}^{\pm} \gamma_{-\vec k}^{\pm}\right] .
\ee  
For the other helicity we would have an overall minus sign. It is straightforward to study the effect of this term in the usual in-in formalism \cite{Maldacena:2002vr}. The correction to the power spectrum is given by
\be
\delta\langle\gamma_{\vec k}^\pm \gamma_{\vec k'}^\pm\rangle = \mp i \int_{-\infty}^\eta \!\!\!\!\rmd \tilde\eta \;\langle \gamma_{\vec k}^\pm(\eta) \gamma_{\vec k'}^\pm(\eta) {\cal H}^{\rm int}(\tilde\eta)\rangle + {\rm c.c.} \;.
\ee
In the late-time limit, $\eta \to 0$, the result does not depend on $\alpha$ and the power spectrum is modified to
\be
\langle\gamma_{\vec k}^\pm \gamma_{\vec k'}^\pm\rangle = (2\pi)^3 \delta(\vec k + \vec k') \frac{H^2}{2\Mpl^2 k^3} \left(1\pm \beta \frac{\pi}{2}\frac{H}{\Lambda}\right) \;.
\ee
The same result was obtained in \cite{Satoh:2010ep}.
For a large background of tensor modes, $r \sim 0.1$, one will be able to distinguish a $50\%$ difference in the power spectra of the two helicities \cite{Gluscevic:2010vv}.

 \vspace{0.1cm}
{\em Enhanced graviton non-Gaussianity? -}
We saw above that it is not possible, at the lowest derivative level, to change the predictions for the power spectrum of tensor modes. We now check that the same happens for the cubic correlator $\langle\gamma\gamma\gamma\rangle$. With three gravitons, the minimum number of derivatives is two.\footnote{In pure de Sitter, i.e.~in the absence of a breaking of time diffs due to the inflaton, this correlator is strongly constrained by the isometry of de Sitter space, so that it can be fixed in terms of three constants, without relying on a derivative expansion \cite{Maldacena:2011nz}. In the presence of the inflaton one cannot get such a general result, but one can rely on the derivative expansion: the correlator will be dominated by operators with the lowest number of derivatives.} If they are both with respect to time, schematically $\dot\gamma\dot\gamma\gamma$, one is forced by invariance under time-dependent spatial diffs to promote $\dot\gamma$ to the extrinsic curvature. The only operator that one can write  is thus $\delta K_{ij}\delta K^{ij}$: as discussed before, this operator does not contain a cubic graviton interaction. It is straightforward to realize that it is impossible to write an operator with one time and one spatial derivative: one may include the totally antisymmetric $\varepsilon$ tensor but cannot build an invariant geometric operator. If the derivatives are both spatial, the operator has only to do with the 3d geometry. The only scalar that one can write  with two derivatives is the 3d Ricci scalar: we saw above this term can always be cast in the standard form inside the 4d Ricci. We conclude that, at two derivative level, the correlator $\langle\gamma\gamma\gamma\rangle$ has always the standard form, first calculated in \cite{Maldacena:2002vr}. Higher derivative corrections start with three derivatives: parity violating operators were discussed above, while parity-conserving ones may have three time derivatives (e.g.~$\delta K_{ij}\delta K_{jl}\delta K_{li}$) or one time derivative (e.g.~$\delta K_{ij} \delta {}^{(3)} \!R$).

It is difficult to reach general conclusions involving mixed correlators. For example, one can induce  an arbitrarily large $\langle\zeta\gamma\gamma\rangle$ with the operators $\delta N \delta K_{ij}\delta K^{ij}$ and $\delta N \delta {}^{(3)} \!R$, though this may be quite unnatural. On the other hand, the $\langle\gamma\zeta\zeta\rangle$ correlator comes, in the standard case, from the tadpole $g^{00}$: it is thus impossible to enhance this correlator, unless one relies on higher-derivative operators.

 \vspace{0.1cm}
{\em Conclusions -}
We showed that the tensor power-spectrum formula $\langle\gamma\gamma\rangle = (H/\Mpl)^2/(2k^3)$, with $H$ and $\Mpl$ Einstein frame quantities, is completely general and only receives (small) higher-derivative corrections. In particular, the tensor amplitude fixes the energy scale of inflation. The tilt of the power spectrum cannot be modified by a time-dependent speed of tensor modes: a blue tensor tilt requires violation of the NEC in the Einstein frame.

\vskip.1cm
\emph{Acknowledgements:} We thank C.~Germani, A.~Maleknejad, M.~Mirbabayi, A.~Moradinezhad-Dizgah, T.~Noumi, D.~Pirtskhalava, S.~Sheikh-Jabbari, L.~Sorbo, M.~Yamaguchi and M.~Zaldarriaga for useful discussions. J.G.~and F.V.~acknowledge financial support from {\em Programme National de Cosmologie and Galaxies} (PNCG) of CNRS/INSU, France and thank ICTP for kind hospitality. J.N. is supported by the Swiss National Science Foundation (SNSF), project ``The non-Gaussian Universe'' (project number: 200021140236).

\appendix



\end{document}